\documentclass[11pt,twoside]{article}

\usepackage{asp2006}
\usepackage{epsf}

\begin{document}

\title{The Formation of Star and Planetary Systems: New Results from {\it Spitzer}}

\author{P. Manoj} 
\affil{Department of Physics and Astronomy, University of Rochester, Rochester, NY 14627, USA}

\begin{abstract}
  Protoplanetary disks are thought to be the birth places of planetary
  systems. The formation and the subsequent evolution of
  protoplanetary disks are regulated by the star formation process,
  which begins with the collapse of a cloud core to form a central
  protostar surrounded by a disk and an overlying envelope. In the
  protostellar phase, most of the envelope material is transferred
  onto the star through the disk during episodic, high accretion
  events. The initial conditions for planet formation in protoplanetary
  disks are likely set by the details of these processes. In this
  contribution, I will review some of the new observational results
  from {\it Spitzer} on protostellar evolution and the structure and
  evolution of protoplanetary disks surrounding young stars in the
  nearby star forming regions. The implications of these results for
  planet formation and eventual disk dissipation are discussed.
\end{abstract}


\section{Introduction}
Stars are formed in the dense and cold cloud cores within
molecular clouds. In the current paradigm for star formation, the
gravitational collapse of a slowly rotating cloud core leads to the
formation of a central protostar surrounded by a rotating disk and an
overlying envelope from which the material rains down onto the
disk \citep{shu87,mo07}. High angular momentum material in the
envelope first collapses into a disk before getting accreted onto the
central protostar. In the early embedded stages, the system drives
powerful bipolar jets/outflows, the origin of which is not entirely
understood. As the system evolves, the envelope dissipates either by
draining onto the disk or is cleared out by stellar winds and outflows,
leaving behind a young pre-main sequence star surrounded by a disk.
Planetary systems are formed out of such protoplanetary disks which
are the natural byproducts of star formation process.

In the widely accepted `core accretion' model, the planet formation
process begins with the sub$\micron$-sized grains in protoplanetary
disks sticking together to grow into larger mm- and cm-sized particles
\citep[e.g.][]{weiden80}. As they grow, the larger grains sink down to
the disk mid-plane; sedimentation of the dust can cause significant
changes in the vertical structure of the disk \citep{dalessio06}.
Along with the grain growth, mineralization of the initially
amorphous dust grains also takes place in protoplanetary disks
\citep[e.g.][]{watson09}. The larger grains that have settled to the
disk mid-plane further grow into km-sized planetesimals, which through
collisional growth, eventually form protoplanets
\citep[e.g][]{weiden08}. Once these `planetary embryos' have become
sufficiently massive ($\sim 10$~M$_{\oplus}$), they accrete gas from
the disks to form giant planets \citep[e.g.][]{pollack96}. A
Jupiter-like gas giant formed in the disk can gravitationally alter
the disk structure by forming radial gaps and holes in them
\citep{rice03,quillen04}. The disks eventually dissipate either by
planet formation processes clearing the disks and/or by various other
disk dispersal processes such as accretion onto the central star,
photoevaporation and magneto-rotational instability (MRI) induced disk
clearing \citep[e.g.][]{alex06I,chiang07}.

\section{Protostellar Evolution}
The various evolutionary stages that young stars go through have
traditionally been classified on the basis of their observed spectral
energy distributions (SEDs). The continuum spectral
index, $\alpha\:\equiv\:d\:log\left(\lambda
  F_{\lambda}\right)/d\:log\left(\lambda\right)$, evaluated between 2
and 20~$\micron$ has been used for this classification
\citep{wilk89,green94}. Under this scheme, objects with a rising
infrared continuum ($\alpha \: \ge \: 0.3$) are classified as Class~I
sources and objects with flat infrared continuum ($-0.3 \: \le \: \alpha 
\: < \: 0.3$) as `Flat spectrum sources'; Class~II objects have infrared
continuum slope $\: -1.6 \:  \le \: \alpha \:  \le \: -0.3\:$ and Class~III
sources show  marginal infrared excess with $\: \alpha \:  < \:  -1.6$. These
empirical SED classes roughly correspond to the different physical
stages in the theoretical model: the Class~I sources have a collapsing
envelope surrounding the protostar-disk system; the Class~II sources
have dissipated most of their envelope and are characterized by a
pre-main sequence star surrounded by a disk; the disk dissipation
is well underway in Class~III sources. An additional class, Class~0,
was introduced later to represent objects in an evolutionary stage
prior to that of Class~I, where the mass in the envelope is
substantially higher than that of the central protostar
\citep{andre93,andremont94}. Traditional SED classes, however, may 
not always represent the actual physical evolutionary stages as
described above, particularly in regions of high extinction: a highly
extinguished Class~II object could easily be misclassified as a  Flat
spectrum or a Class~I source \citep{mcclure10, evans09}.

The number of objects in various SED classes have been used to
estimate the relative lifetimes of the evolutionary stages
\citep[e.g.][]{wilk89,green94,kenhart95}. Such studies carried out
before the {\it Spitzer} era yielded an average lifetime of $\la
0.1$~Myr for the Class~0 phase \citep{andre93} and $0.1 - 0.2$~Myr for
Class~I phase \citep{green94,kenhart95}. Many of these studies
suffered from small number statistics and possible differences in the
earlier evolutionary classes between clouds. The large {\it Spitzer}
survey of five nearby molecular clouds by the {\it c2d} team have
significantly improved the data statistics \citep{evans09,enoch09}.
The lifetimes of the protostellar evolutionary classes are estimated
from the number counts of objects in various SED classes, under the
assumptions of continuous star formation and a mean lifetime of 2~Myr
for the Class~II objects.  The mean lifetime for the Class~I phase was
found to be $0.4 - 0.6$~Myr with an additional  0.4~Myr for the
Flat SED phase \citep{evans09}.  The average lifetime derived for the
Class~0 phase is $0.1-0.2$~Myr.  These lifetimes are
significantly longer than most of the earlier estimates in the
literature \citep{evans09}.

\subsection{The `Luminosity Problem'}

Most of the stellar mass is built during the embedded protostellar
phase which lasts for $\sim 0.5$~Myr \citep{evans09}. A steady
accretion rate of $1.0 \: \times \: 10^{-6}$ M$_{\odot}$~yr$^{-1}$
would then be required to build a 0.5~M$_{\odot}$ star, which
corresponds to an accretion luminosity of L$_{acc}$ = 6~L$_{\odot}$.
However, the observed median bolometric luminosity of most of the
Class~I protostars in the nearby clouds is  $\la 1$~L$_{\odot}$
\citep{ken90,motte01}, significantly smaller than the expected
accretion luminosity. Moreover, the observed luminosity is an upper
limit for L$_{acc}$: only a small fraction of L$_{bol}$ is
attributable to accretion \citep{muz98,wh04}. This `luminosity
problem' of protostars, first recognized by \citet{ken90}, is
essentially an accretion rate problem: the observed accretion rates in
Class~I protostars are significantly smaller than those required to
build a star during the embedded protostellar phase.

The {\it Spitzer} observations of protostars in the nearby clouds
confirm the `luminosity problem' \citep{evans09,enoch09}. The observed
L$_{bol}$ of most Class~I sources are significantly lower than those
predicted by the standard infall models with steady accretion
\citep{evans09, dunham09}. The L$_{bol}$ distribution shows a
dispersion of several orders of magnitude, with a few objects
displaying higher L$_{bol}$ values than predicted by the models.
Furthermore, {\it Spitzer} observations do not show any clear evidence
for an early phase of rapid mass accretion, suggesting that most of
the stellar mass is built during the Class~I phase \citep{evans09}.
The observed distribution of L$_{bol}$ of the Class~I protostars in
nearby clouds is better explained by protostellar evolutionary models
with episodic mass accretion, rather than those with steady accretion
\citep{dunham09}. All the available evidence seem to indicate that the
accretion onto the protostar during the embedded phase is non-steady 
and time variable.  This was first proposed by \cite{ken90}, who
suggested that the accretion during the embedded phase must be
episodic with prolonged periods of very low accretion. Most the of the
stellar mass is built during the high accretion events.

It has been suggested that the high accretion stage of the
protostellar phase corresponds to the FU~Ori objects \citep{ken90,
  hart98}. FU~Ori systems exhibit sudden increases in brightness
($\Delta V \sim 5$~mag) and remain luminous or decay slowly over
decades \citep{herbig77,hartken96}. The FU~Ori outbursts are believed
to be caused by the rapid increase in the accretion rates which can be
as high as $10^{-4}$~M$_{\odot}$~yr$^{-1}$ during the outburst
\citep{hartken96}. In the low accretion state of the protostars, the
material falling in from the envelope onto the disk at a much higher
rate, piles up in the outer disk. The accumulation of the material
makes the disks unstable; the onset of thermal or gravitational
instabilities in the disk then transfers the bulk of the disk material
onto the central star at high accretion rates,  which is  observed as the
brief high luminosity FU~Ori outburst \citep{hartken96,hart98}.

\begin{figure}[!ht]
\plottwo{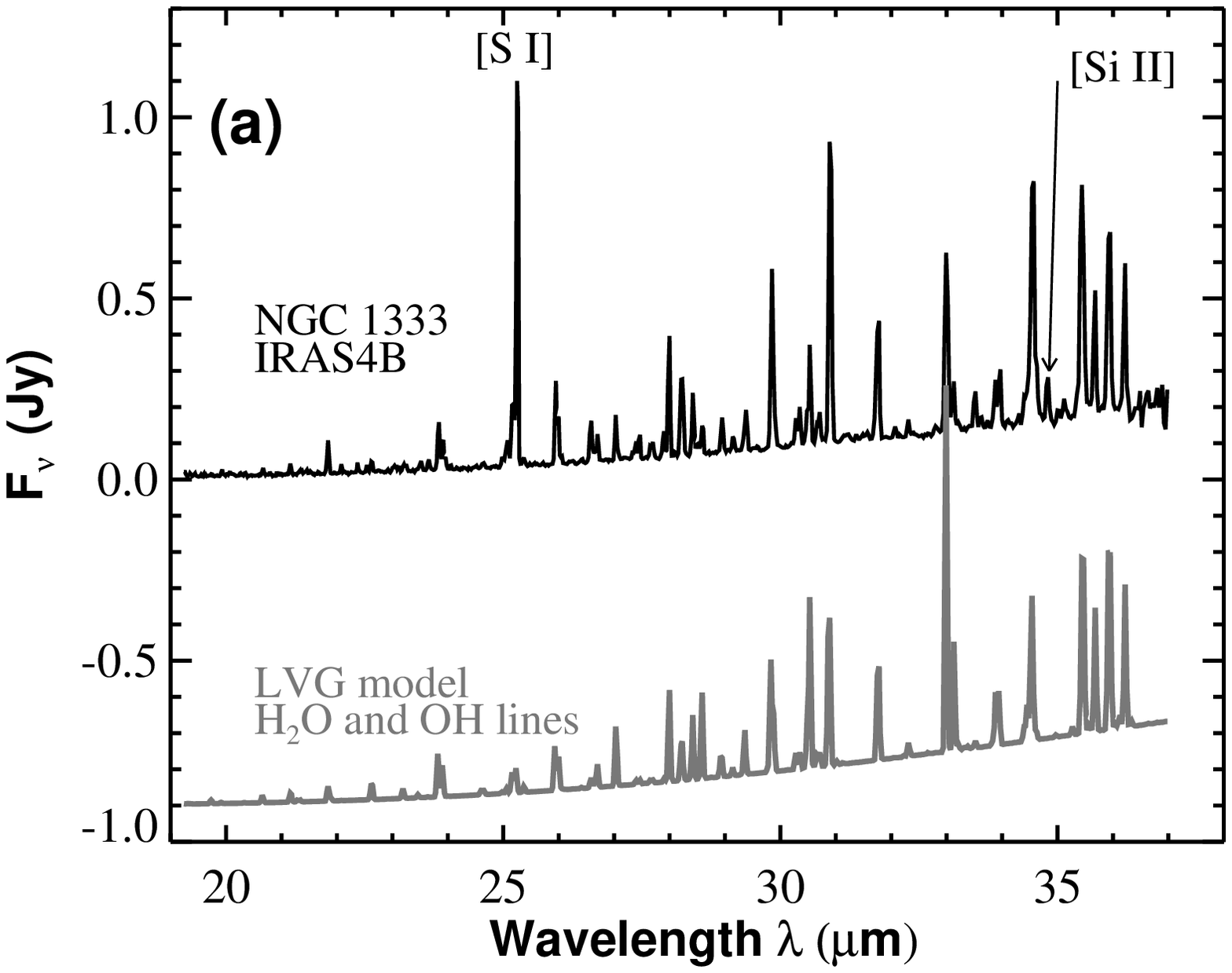}{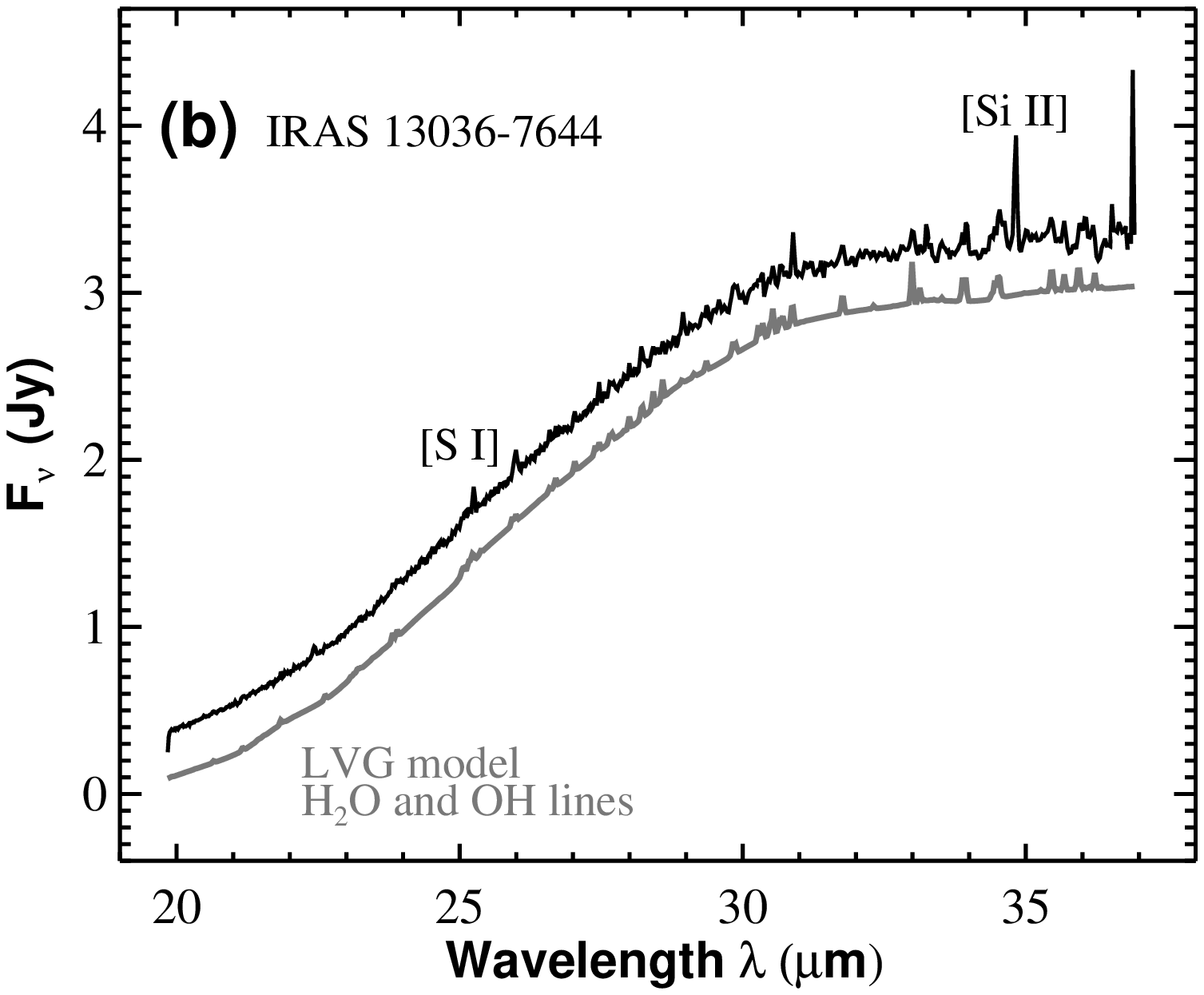}
\caption{High-resolution {\it Spitzer}-IRS spectra (black line) of
{\it (a)} NGC 1333 IRAS~4B \citep{watson10} and {\it (b)}
IRAS 13036-7644 \citep{manoj10b}. Except the [S I] and
[Si II] lines, which originate from outflows, all the other lines seen
in the spectra are from pure rotational transitions of water and
OH. In both the panels, the observed spectra are compared to a simple
plane-parallel LVG model (gray line,) showing water and OH lines, which
originate in an envelope-disk accretion shock
\citep[see][]{watson07}. \label{fig1}}
\end{figure}

\subsection{Envelope-Disk Accretion in Protostars}
The envelope infall signatures in Class~0/Class~I protostars have been
observed in the molecular lines in the submm and mm wavelengths
\citep{ohashi04}.  However, these observations typically trace spatial
scales of $\sim 0.01$~pc and densities of $10^5 - 10^6$~cm$^{-3}$ and
thus probe large scale infall motions in the envelopes
\citep[e.g.][]{difranc01}.  Direct observational evidence for the
material falling onto the embedded disks in a protostar has been
lacking until recently. The freely falling envelope material which
acquires supersonic terminal infall velocities is thought to be
decelerated within the accretion shocks at the disk surface
\citep{ulrich76,cm81,nh94}. One of the most important {\it Spitzer}
discoveries in protostellar studies has been the detection of the
arrival of infalling envelope material on the disk in an envelope-disk
accretion shock, through the mid-infrared emission lines of water and
OH lines observed in protostars viewed face-on with the Infrared
Spectrograph (IRS). This was first detected in NGC~1333-IRAS~4B, which
shows a rich emission-line spectrum of water and OH \citep{watson07}
(see Figure~\ref{fig1}). The relative intensities of the water lines
indicate their origin in extremely dense ($10^{10}-10^{12}$~cm$^{-3}$)
and warm ($\sim 170$~K) gas.  Detailed modeling shows that the gas is
heated by a low velocity ($\sim 2$~km~s$^{-1}$) envelope-disk
accretion shock, caused by the envelope material landing on the disk
in an annulus of $40-60$~AU.  The rate at which the envelope material is
falling on the disk, obtained from the total water line luminosity
under the assumption of water line emission being the major coolant,
is found to be $0.7 \: \times \: 10^{-4}$~M$_{\odot}$ yr$^{-1}$, quite
similar to the envelope infall rates estimated from molecular line
observations \citep[for details see][]{watson07}.

\begin{figure}[!ht] 
\plotone{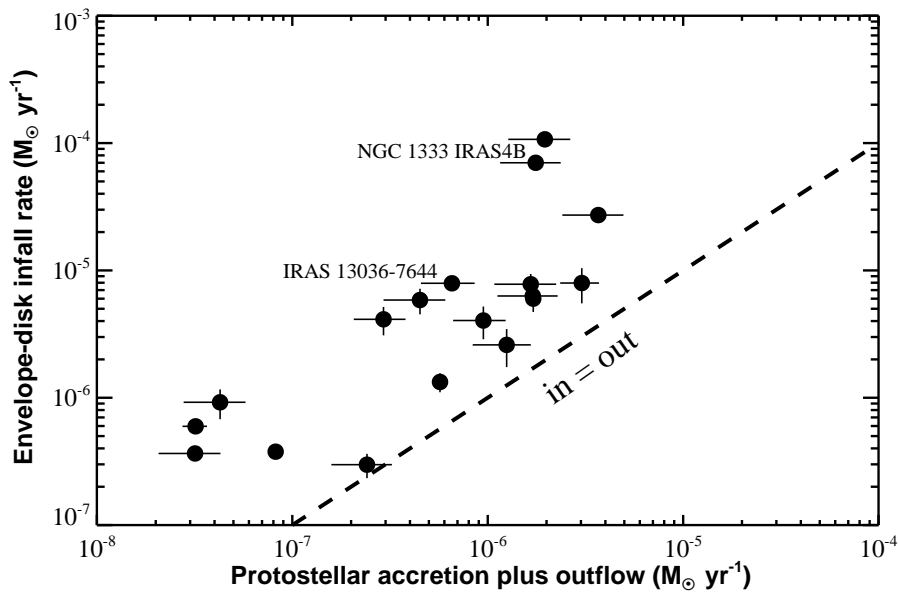}
\caption{The envelope-disk accretion rate, estimated from the water and
OH lines in the mid-IR spectra, of 18 protostars plotted against the
protostellar accretion plus outflow rate. The dashed line represents infall rate
equal to the protostellar accretion plus outflow rate. IRAS 4B and
IRAS 13036-7644, whose spectra are shown in Figure~\ref{fig1}, are
labeled. Most protostars observed, show significantly higher envelope
infall rates compared to the protostellar accretion and outflow rates,
indicating that the material is accumulating in their disks.
\label{fig2}}
\end{figure}

\citet{watson10} carried out a survey of 85 face-on Class~0/Class~I
protostars within 500~pc, with the {\it Spitzer} IRS high resolution
modules to look for signatures of envelope-disk accretion and outflows
in them. They detected water and OH lines in roughly half of their
sample. An example of this, the high resolution mid-IR spectrum of
IRAS~13036-7644, which is a bright Class~0 source in the small cloud
DC303.8-14.2 towards Chamaeleon II, is shown in Figure~\ref{fig1}
\citep{manoj10b}. The modeling of the spectra of the protostars, which
show mid-IR emission lines of water and OH, indicates that the
conditions in the envelope-disk accretion shock in them are similar to
that found for IRAS4B.  With the new diagnostic of envelope-disk
accretion rate made available by {\it Spitzer}, it is now possible to
compare the various flows in a protostellar system. In
Figure~\ref{fig2} the envelope-disk infall rate of the observed
protostars is compared to the protostellar accretion plus outflow
rate. The envelope-disk infall rate is estimated from the total
luminosity of the water lines observed \citep[see][]{watson07}. The
protostellar accretion rate is computed from L$_{bol}$, assuming
$\frac{R_{\star}}{M_{\star}}$~=~$\frac{10\;R_{\odot}}{M_{\odot}}$,
following the prescription of \citet{evans09} and \citet{enoch09}. The
outflow rate is computed from the luminosity of [Si~II] line at 34.8
$\micron$ in the IRS spectra. The [Si~II] line is a good proxy for the
[O~I] 63.2 $\micron$ line which measures the outflow rates, as it is a
dominant coolant for densities of $10^{3}-10^{4}$~cm$^{-3}$
and outflow speeds of $30-150$~km~s$^{-1}$ \citep{hm89}. It is
clear from Figure~\ref{fig2} that the envelope infall rates derived
from water and OH lines for most protostars are at least an order of
magnitude higher than the protostellar accretion and the outflow rate.
This indicates that material is piling up in the disk and is not
efficiently accreted onto the star or carried out by the outflow. This
could eventually lead to FU~Ori-like outburst which is caused when the
material that is accumalated in the disk is accreted onto the central
protostar in a massive accretion event.  Thus, the flow imbalance in
protostars, measured from their mid-IR spectra, is consistent with,
and argues strongly for episodic accretion in protostars.

\section{Structure and Evolution of Protoplanetary Disks}

The embedded phase in the protostellar evolution ends after most of
the envelope mass has been transferred to the central star through the
disk. A pre-main sequence star surrounded by an accretion disk is left
behind. In such Class~II objects, the emission from protoplanetary disks
dominates the SED longward of 2 $\micron$. The thermal emission from
the dust grains in the disks, heated by the stellar radiation, is
responsible for the excess emission in these young stars. The
optically thin emission from the dust grains in the surface layers of the
disk produces the silicate emission features centered at 10 and 20
$\micron$ seen in the mid-IR spectra. The optically thick emission
from the deeper layers and the disk midplane dominates the mid-IR
continuum. The detailed shape of the mid-IR continuum is related to
the disk geometry and could be used to probe the vertical and radial
structure of the disks \citep{dalessio06}. The silicate emission
features inform us on the size, shape and composition of dust grains
in the disks \citep[e.g.][]{bouwman01}.  {\it Spitzer} observations in
the mid-IR, probe the planet forming regions ($\sim 1-10$~AU) of
protoplanetary disks \citep{dullemond07,dalessio06} and thus provide
powerful diagnostics for the structure and evolution and the dust
mineralogy of these disks.
 
Most solar-mass stars younger than $\sim 1$~Myr appear to harbor disks
around them; by $\sim 3-5$~Myr most of them shed their disks
\citep{haisch01,hill05,hernandez08}. Clues to the planet formation and
disk dissipation processes must therefore be sought in $1-2$~Myr
protoplanetary disks where disk evolution in `action' could be
studied. Below, I summarize the results from the large {\it Spitzer}
IRS survey of protoplanetary disks in the nearby  $1-2$~Myr  old star forming
regions of Taurus, Chamaeleon I (Cha I) and Ophiuchus carried out by the
{\it IRS\_disks} team (PI: Dan M. Watson).

\subsection{Dust Sedimentation in Disks}
The first step towards planet formation is the growth of
sub$\micron$-sized grains in protoplanetary disks via coagulation.  As
the grains grow larger, they sink towards the disk mid-plane; both the
grain growth and sedimentation deplete the amount of small dust
grains in the disk surface layers. This reduces the opacity offered by the
disk surface layers to the stellar radiation and as a result the
amount of energy absorbed by the surface layers decreases.
Consequently, the degree of flaring of the disk decreases, making the
disks flatter: higher the degree of sedimentation, flatter are the
disks \citep{dalessio06, dullemond07}.  The changes in the disk
geometry is readily reflected in the mid-IR SEDs. Self-consistent
accretion disks models, which incorporate the effects of dust settling,
show that the shape of the mid-IR continuum is a good measure of the
degree of dust settling \citep{dalessio06}. In particular, the SED
slope between 13 and 31 $\micron$ has been used to
characterize the vertical distribution of the dust in the disks
\citep{dalessio06,furlan06,watson09}.

\begin{figure}[!t] 
\plotone{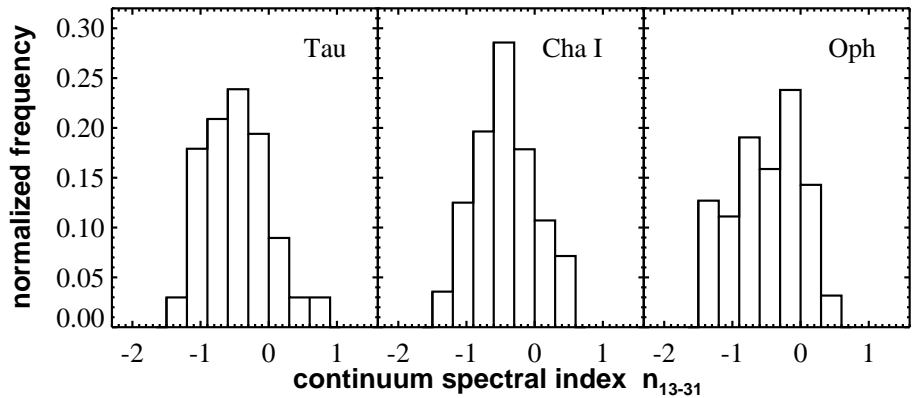}
\caption{The distribution of mid-IR continuum spectral slope between
13 and 31 $\micron$ evaluated from the dereddened {\it Spitzer} IRS spectra for
Class~II objects in Taurus, Chamaeleon I and Ophiuchus
regions \citep[also see][]{furlan06, mcclure10, manoj10}. \label{fig3}} 
\end{figure}

A careful comparison of the observed mid-IR continuum slope of young
protoplanetary disks in Taurus, Cha I and Ophiuchus with that
predicted by the models shows that significant dust settling has
occurred in $1-2$~Myr old protoplanetary disks. The dust-to-gas mass
ratio in the surface layers of most of these disks is found to be
lower by factors of  100 to 1000 compared to the standard ISM value,
indicating a high degree of sedimentation in these disks
\citep{furlan06, mcclure10, manoj10}.  Further, the degree of
sedimentation and the frequency of sedimented disks in the $\sim 1$~Myr old
Taurus, the $\sim 2$~Myr old Cha I and the $\la 1$~Myr Ophiuchus star forming
regions are found to be similar \citep{furlan09, mcclure10, manoj10}.
This is demonstrated in Figure~\ref{fig3}, where the distribution of
the continuum spectral slope evaluated between 13 and 31 $\micron$
from the {\it Spitzer} IRS spectra of the young stars in these three
regions are compared. The range and frequency of the distribution of
the observed n$_{13-31}$ index in quite similar in all the three
regions.  This suggests that significant dust settling occurs in
protoplanetary disks by $\la 1$~Myr. A comparative study of the median
mid-IR SEDs of these three regions also confirms that overall, the
vertical structure of the disks in these three regions are very
similar \citep{furlan09}.

\subsection{Grain Growth and Crystallization}
The silicate emission features observed in the mid-IR spectra of young
stars have been used to infer the dust properties in protoplanetary
disks. Silicate emission feature produced by the sub$\micron$-sized
amorphous grains in the disks tend to be narrow and peaked. Grain
growth and crystallization makes the silicate emission profile broad
and structured \citep[e.g.][]{bouwman01}.  Detailed analysis of the
shape and strength of silicate features in the {\it Spitzer} IRS
spectra of a large sample of Class~II objects in the Taurus region,
show evidence for significant grain growth and crystallization in
those disks \citep{watson09}. More quantitative spectral decomposition
modeling of the IRS spectra shows a wide range in the degree of grain
growth and crystallization in young  disks \citep[e.g.][]{sargent09,
  juhasz09}. In general, the degree of dust processing (grain growth
and crystallization) is found to correlate well with the degree of
dust sedimentation in disks: less sedimented disks show evidence for
relatively less processed dust whereas flat and highly sedimented
disks appear to have highly processed grains \citep{watson09,
  sargent09, mcclure10}.  However, no strong correlations have been
found between various stellar parameters and the degree of dust
processing \citep{watson09}.

\subsection{Transitional Disks}
\begin{figure}[!t]
\plottwo{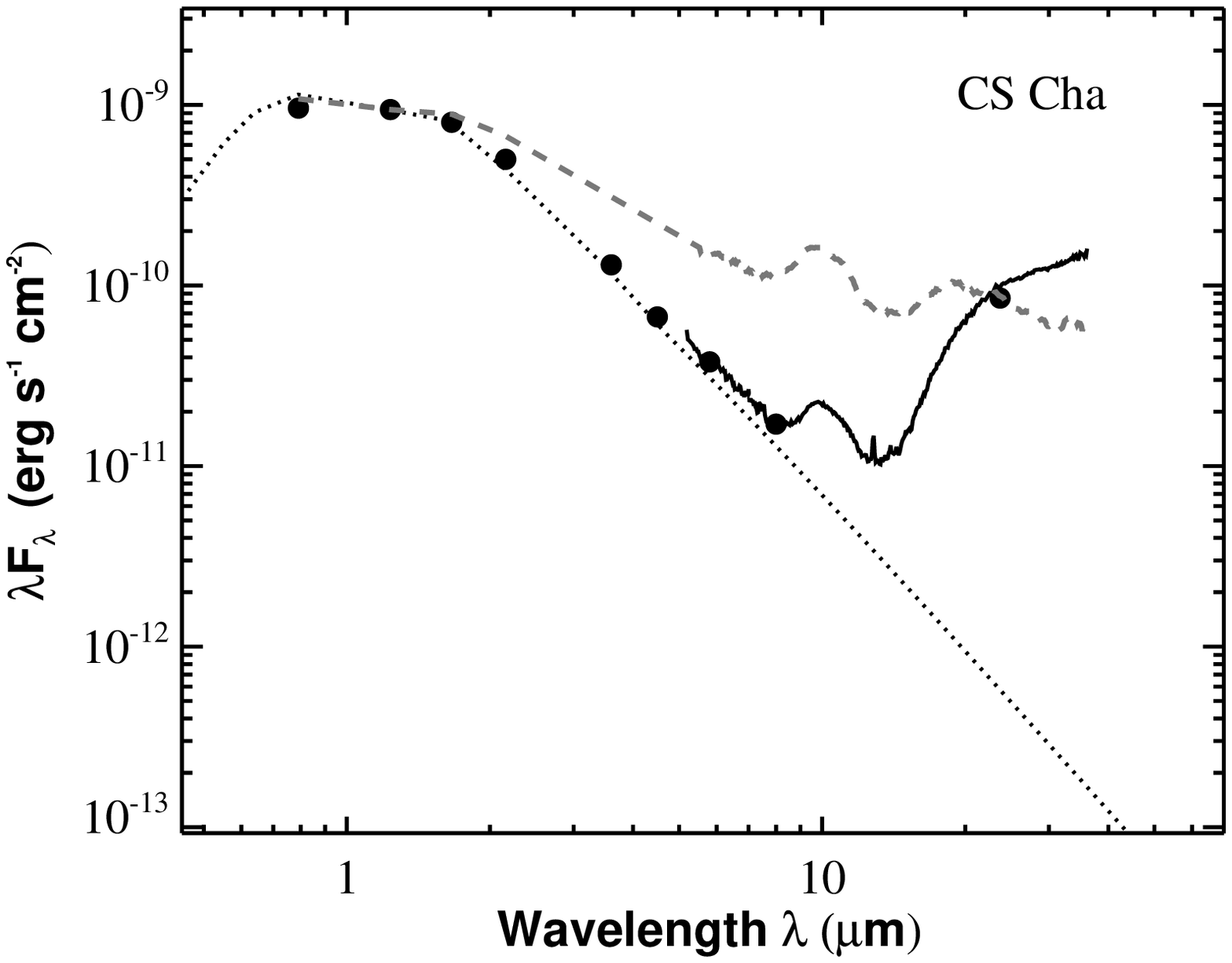}{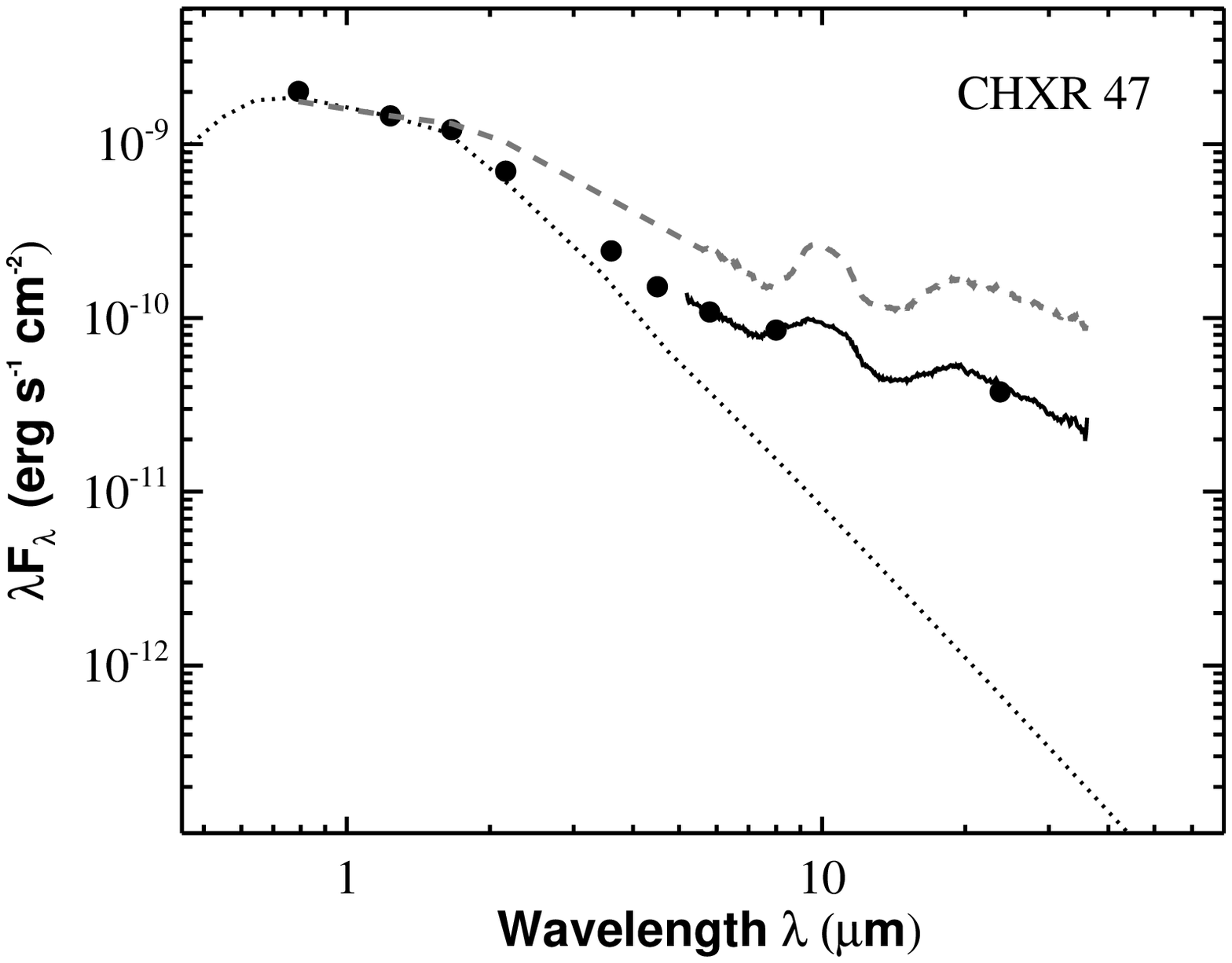}
\caption{Examples of the observed SEDs of a transitional disk ({\it left}) and an evolved disk ({\it right}) in the Cha~I star forming region. The solid symbols represent I, J, H, K and {\it Spitzer} IRAC and MIPS 24 $\micron$ photometry and the black solid line the {\it Spitzer} IRS spectrum. The gray dashed line represents  the median SED of Class~II objects in Cha~I. \label{fig4}}
\end{figure}

One of the important contribution of {\it Spitzer} in the area of disk
evolution has been the identification and characterization of a large
number disks with significantly altered radial structures. Such disks
have been called `transitional disks' to suggest that they are
possibly in an evolutionary stage in between that of an optically
thick disk and an optically thin disk \citep{skrut90}. Their SEDs are
characterized by a significant deficit of flux at wavelengths $\la
10$~$\micron$ compared to that of the optically thick, full disks,
along with a sharply rising continuum and excess emission comparable
to, or higher than, that of a typical Class~II disk at mid- and
far-infrared wavelengths \citep{calvet05,esp07a,kim09,furlan09} (see
Figure~\ref{fig4}).  Detailed modeling of their SEDs shows that
transitional disks have inner holes or gaps in the radial dust opacity
distribution that are surrounded by an optically thick outer disk with
a `wall' at its inner edge. Emission from this `wall', where the
abrupt transition from the inner optically thin region to outer
optically thick disk takes place, is responsible for the sharply
rising continuum longward of 13~$\micron$ observed in these objects
\citep{calvet05, esp07b, esp08b}.  Several transitional disks have now
been identified from {\it Spitzer} photometric and spectroscopic
surveys of nearby young clusters \citep{furlan09, kim09, muzerolle10}.
Most transitional disks are Classical T Tauri stars (CTTs) meaning
that they are active accretors; however, their accretion rates are
slightly lower than the median accretion rate of other Class~II disks
in the region \citep{najita07}.

The presence of holes and gaps in transitional disks, inferred from
their observed SEDs, suggests that significant disk evolution has
occurred in these disks. Various disk evolutionary mechanisms have
been proposed to explain the origin of inner holes/gaps in
transitional disks: substantial grain growth, planet formation,
photoevaporation and MRI-induced inner disk clearing are some of them
\citep{dd05, mm92, quillen04, alex06I, chiang07}. It has been
suggested that the Jovian mass planets opening up gaps and holes in
the disks can explain most of the observed characteristics of a large
sample of transitional disks \citep{kim09, najita07, rice03,calvet05,
  esp07b}.

In close binary systems (few AU separation) stellar companions can
clear inner holes in the circumbinary disks \citep{artlub94}; the
observed SEDs of such systems would appear transitional-disk-like, as
has been demonstrated in the case of CoKu Tau/4 \citep{ik08}. However,
as argued by \citet{muzerolle10}, not all close binaries show evidence for 
significant inner holes. Objects which show these SED characteristics
are likely to be in transition.

\subsubsection{Evolved Disks}
The large photometric and spectroscopic surveys with {\it Spitzer}
have also identified another class of disks whose excess emission is
significantly lower than that of the Class~II median for that region.
These disks differ from the transitional disks discussed above in that
their excess flux, even at longer wavelengths, is considerably lower
than that of the Class~II median. Also, they do not have a rising
continuum longward of 13~$\micron$ indicating that the outer disk
probably is not as optically thick as that of Class~II disks and the
`wall' emission is not significant. This class of disks has been
identified by various names in the literature: `evolved disks',
`anemic disks', `homologously depleted disks' \citep{lada06,
  hernandez07, currie09a, luhman09}. The SED of an evolved disk is
compared to that of a `transitional disk' in Figure 4.

Another crucial difference between classical transitional disks which
have inner holes/gaps and the evolved disks is that the most evolved
disks are Weak-line T Tauri stars (WTTs). They have stopped accreting
or are accreting at very low rates \citep{muzerolle10, manoj10}. It is
not clear if evolved disks fit into an evolutionary sequence in which
the optically thick disk develops inner gaps and holes and eventually
dissipates inside out. It is possible that they represent a different
evolutionary path for the primordial optically thick disks where the
disk dissipation does not proceed inside out. In this sense, evolved
disks have also been taken to be in a transitional phase between a
primordial optically thick disks and optically thin debris disks
\citep{lada06, currie09a, muzerolle10}. However, it is unclear what
mechanisms can dissipate primordial disks simultaneously at all radii.

\subsection{Disk Lifetimes}
Ground based surveys which used the near-IR excess and emission lines
diagnostics to probe the innermost parts ($\ll 1$~AU) of the disks
have shown that the median lifetime of accretion disks around young
stars is only a few Myr \citep{haisch01,hill05,hernand05,manoj06}.
{\it Spitzer} observations in the mid-IR, which probe the planet
forming regions in the disk ($\la 10$~AU), have confirmed this. The
{\it Spitzer} studies of disk fraction in clusters show that most
disks dissipate in about $3-5$~Myr \citep[e.g.][]{hernandez08,
  mamajek09}.  A small fraction of disks appear to last longer: the
well known examples of long-lived accretion disks include members of
$\eta$~Cha~(6~Myr), TW~Hya~(8~Myr) and 25~Ori~(8~Myr) groups, PDS 66
($7-17$~Myr) and St~H$\alpha$~34~($8-25$~Myr) \citep[][and references
therein]{mamajek09}.  Disk dissipation timescale is also found to
depend on the central stellar mass: disk lifetimes are shorter for
intermediate mass (spectral type late B to F) stars \citep{carp06,
  lada06, hernand07b,currie09a} and marginally longer for brown dwarfs
compared to that for solar-mass (spectral type K to early M) stars
\citep{luhman05, mamajek09}.

The observed low frequency ($\la 10$\%) of transitional disks in young
clusters, has long been taken as evidence for a relatively short-lived
(less than a few hundred thousand year) transitional phase, during
which an optically thick disk turns into an optically thin disk
\citep{skrut90, ww96, pratosimon97}. The primordial disks last for a
few Myr, but disk dissipation, once initiated, occurs rapidly, on
timescales $\ll 1$~Myr. This is sometimes referred to as the
`two-timescale' behavior of disk dissipation \citep{alex06I}.  The
{\it Spitzer} studies have significantly improved the statistics of
well characterized transitional disks in nearby young clusters.
Some studies have reported a relatively high fraction ($40-50$\%) of
transitional and evolved disks in a few young clusters
\citep{sicilia08, currie09a}.  Based on this, it has been argued that
the lifetimes of transitional disks are comparable to that of
primordial disks and that the disk dissipation is not as rapid as
previously thought \citep{currie09a, currie09b}. However, in some of
these studies, flat well-settled disks and primordial disks around
late M type stars have been misidentified as transitional or evolved
disks, thereby increasing their apparent frequency
\citep[see][]{luhman09, furlan09, ercolano09}. The frequency of
transitional and evolved disks has also been found to have an age
dependence: a few percent at $1-2$~Myr rising to $10-20$\% at
$3-10$~Myr \citep{muzerolle10}.  While this trend may be real, the
higher frequency of disks in transition in older clusters may not
necessarily imply a longer transition timescale. It is more likely
that star formation has stopped in older ($\ga 3$~Myr) clusters, and
as a result, there is no steady supply of primordial disks, which
shoots up the ratio of the number of transitional disks to primordial
disks.  When the different star formation histories of $2-10$~Myr
clusters are properly taken into account, the transitional phase
appear to last only for about $10-15$\% of the median primordial disk
lifetime ($\sim 3$~Myr) \citep{luhman09}.

\section{Dispersion in Disk Evolution}
The evolution of protoplanetary disks has now been studied using
various observational diagnostics, including the disk properties
measured in the mid-IR wavelengths by {\it Spitzer}. The disk
evolutionary indicators such as disk fraction in a cluster, disk
optical depth, accretion rate, disk masses, degree of dust
sedimentation and grain processing in the disk, on average, show a
discernible evolutionary trend with age which is broadly consistent
with the theoretical expectations \citep[e.g.][]{hernandez08,
  furlan09, hart98, carp05}.  However, the observed disk properties also
show a large dispersion. The disk fraction in clusters, in general,
decreases with age, but in every age bin there is a large scatter
\citep{hill05, hillen08, mamajek09}.  While most disks dissipate in
about $3-5$~Myr, a significant fraction ($20-30$\%) of stars appear to
lose their disks as early as $\sim 1$~Myr \citep[see][]{hernandez08, mamajek09}.
A small fraction of stars retain their disks at ages $\ga 5$~Myr.
Even in a given $1-2$~Myr old cluster, disk optical depths (measured as
the IR slope and/or IR excess) show a large dispersion. While most
disks show evidence for significant dust settling and are relatively
flat, a few disks with a high degree of flaring show no evidence for
dust sedimentation \citep{dalessio06, furlan06, watson09, mcclure10,
  manoj10}.  Transitional disks have developed radial holes/gaps in
them while the evolved disks are about to become optically thin at all
radii \citep{kim09, furlan09, mcclure10, manoj10}. The measured
accretion rates of young stars in a cluster also show a large
dispersion: there are high accretors (accreting at rates of $\sim 10^{-7}$~M$_{\odot}$~yr$^{-1}$) and non-accretors among systems with
similar central star properties \citep{hart98b}. Disk masses of
otherwise identical systems in the same  cluster are different by a few
orders of magnitude \citep{aw05}.  The degree of grain growth and
crystallization inferred for T Tauri disks in a cluster span a
wide range \citep{watson09, sargent09, olof09}.

Is the observed dispersion in disk properties in a cluster, caused by
the age spread within the cluster?  It is unclear if the
dispersion in the measured luminosity distribution of members of a
cluster or association indicate a true age spread or observational or
other random errors. Some authors have suggested that the WTTS and
transitional disks are older than CTTS in a cluster
\citep{bert07,fang09} while others find no evidence for such an age
difference \citep{kenhart95,herbig98,hart01,hd02,ds05}.  True age
spreads are likely to be much smaller than the apparent age spreads
inferred from the luminosity dispersion \citep{hart01, hillen08b}.
Recent Monte carlo simulations suggest that bulk of the cluster
members are consistent within errors as having the same age with the
exception of a few outliers \citep{hillen08,hillen08b,hillen09}.  It
appears that the observed dispersion in disk properties in a cluster
cannot be entirely due to the age spread. The stars of apparently same
ages can have different disks. Disk evolution probably is not a
function of age alone. Other possible factors contributing to the
observed dispersion in disk properties are stellar multiplicity
\citep[see][]{bouwman06,cieza09} and effect of immediate environment
such as cluster density \citep{luhman08} and proximity to an O-type
star \citep{balog06,hernandez08, mercer09}. However, it is unclear if
these effects can fully explain the observed dispersion in disk
evolution. Current evidence seems to suggest that systems with similar
stellar mass, age, multiplicity status and environment can have
different disk properties \citep[e.g.][]{watson09, furlan09,
  mcclure10, manoj10}.

How else do we understand the observed dispersion in disk evolution in
otherwise identical systems?  If all the disks started out the same -
i.e. if the initial conditions like disk mass, surface density and
radius are the same - then is the observed dispersion a result of
different disk dissipation mechanisms, with different characteristic
timescales being dominant in different disks? Or is it that these
mechanisms start operating at different times in different disks?
Could it be that the disk dissipation and planet formation processes
are stochastic in nature?  Indeed, planet formation simulations show
that disks with similar initial properties can result in very
different final configurations \citep{thommes08}.

However, if the initial configurations of protoplanetary disks are
different, they would look different at any instant of time, even if
they all go through the same temporal evolutionary sequence.  Some of
the self-consistent simulations of planet formation in protoplanetary
disks show that the outcome of the process strongly depends on the
initial conditions of the disks such as disk mass and the viscosity
parameter $\alpha$, among others. Disks with an initial mass of
0.1~M$_{\odot}$ and low $\alpha$ ($\sim 0.001$) tend to form giant
planets early, which migrate extensively and clear out the disks
faster. On the other hand, low mass disks ($\la 0.01$~M$_{\odot}$)
with high $\alpha$ ($\sim 0.01$) are too slow to form planets and last
longer \citep{thommes08}. In this picture, some of the long-lived
disks such as TW~Hya, PDS~66 and St~34 could  represent disks
which have failed to form planets \citep[see][]{hart05b}, the
controversial detection of a hot Jupiter around TW~Hya notwithstanding
\citep{set08, huel08}. Thus, the later evolution and the appearance of
a few Myr old protoplanetary disks are affected and shaped by their
initial conditions. It is quite possible that the observed dispersion
in the properties of disks in a cluster is just a reflection of the
dispersion in the initial properties of these disks.

The initial conditions during the formation of protoplanetary disks is
set by the star formation process. If episodic accretion is the norm
during the embedded phase of the protostellar evolution, then the
disks undergo a series of high accretion events. The frequency and
magnitude of these events must clearly have an effect on the
subsequent evolution and the observed properties of protoplanetary
disks. The observed dispersion in the disk evolution could then partly
be understood as the star formation process determining the initial
disk configuration before the onset of planet formation.

\acknowledgements I thank the organizers of the Bash Symposium for
giving me the opportunity to participate. Thanks to Dan M.  Watson,
Joel D. Green and other members of the {\it IRS\_disks} team for 
helpful comments.

\end{document}